# Enhancing the electron mobility via delta-doping in SrTiO$_3$

Y. Kozuka,[1,a)] M. Kim,[1] H. Ohta,[2,3] Y. Hikita,[1] C. Bell,[1,3] H. Y. Hwang[1,3,4]

[1]*Department of Advanced Materials Science, University of Tokyo, Kashiwa, Chiba 277-8561, Japan*

[2]*Graduate School of Engineering, Nagoya University, Furocho, Chikusa, Nagoya 464-8603, Japan*

[3]*Japan Science and Technology Agency, Kawaguchi, Saitama 332-0012, Japan*

[4]*Department of Applied Physics and SLAC Photon Science, Stanford University, Stanford, CA 94305, USA*

## Abstract

We fabricated high-mobility δ-doped structures in SrTiO$_3$ thin films in order to investigate the low temperature electronic transport properties of confined carriers in this system. An enhancement of the electron mobility above the bulk value was observed as the doped layer thickness decreased. High-field Hall measurements revealed that this mobility enhancement originates from higher-mobility electrons in the undoped clean regions, which have quantum-mechanically broadened from the doped layer. Because of the absence of apparent lattice misfit between the layers, this structure is highly suitable for investigating two-dimensional electron gases in SrTiO$_3$.

[a)]Present address: Institute for Materials Research, Tohoku University, Sendai 980-8577, Japan; electronic mail: kozuka@imr.tohoku.ac.jp

The artificial engineering of oxide heterostructures has given rise to a wealth of fascinating physical properties, which have been intensively developed in the last decade. Among the many oxide materials, the behavior of low-dimensional carriers in $SrTiO_3$ has been focused on because of its versatile functionality, such as room-temperature blue-light emission,[1] enhanced thermoelectric efficiency,[2] and interface metallic conductivity.[3,4] Additionally, two-dimensional (2D) high-mobility conducting layers of $SrTiO_3$ may form a variety of quantum ground states at low temperatures.[5-8] These properties of $SrTiO_3$ are fascinating in that its large dielectric constant enhances the mobility in a $d$ electron system,[9,10] whereas much of the interest in the physics of high-mobility 2D electron gases has so far been confined to studies in $sp$ hybrid systems.

Despite such promising potential, however, the degree of carrier doping control in $SrTiO_3$ structures is still limited. At the $LaAlO_3/SrTiO_3$ interface, for example, the transport properties are dramatically changed by subtle variation of growth conditions such as oxygen partial pressure[8] and $LaAlO_3$ thickness.[11] Although the transport properties can be tuned using the electrostatic field effect,[11] the carrier distribution is concurrently altered by external electric field, which is accelerated by the electric-field dependent dielectric constant of $SrTiO_3$.[12-14] To draw fully on the potential of these fascinating materials, both the carrier concentration and distribution must be independently well controlled to a degree approaching the levels of current semiconductor technology, which has developed thanks to longstanding efforts to improve the growth and doping techniques of thin films.[15] In the case of $SrTiO_3$, the growth of high-quality, uniformly doped, thin films has recently been reported.[16,17] Based on these growth methods, continued progress is essential to form controlled two-dimensional structures, enabling the realization of novel low-dimensional states in SrTiO$_3$.

Here, we show that δ-doping can confine electrons in a thin layer and lead to significant mobility enhancements beyond bulk values. The structure is depicted in Fig. 1(a), where a Nb-doped SrTiO$_3$ layer is embedded between undoped SrTiO$_3$ buffer and cap layers. In this structure, lattice misfit at the interfaces is virtually absent, which avoids severe degradation of the film quality.[18] As a consequence of this design, electrical transport measurements showed an enhancement of the low-temperature electron mobility with decreasing dopant layer thickness (*d*). The carrier distribution around the doped layer was also estimated by thermoelectric measurements, which confirmed narrow electron confinement by the dopant layer. We conclude that the carrier distribution in this structure leads to a mobility enhancement in analogy with δ-doped systems in conventional semiconductors.[19] These results demonstrate that δ-doped structures can provide a prototypical foundation to develop high-mobility low-dimensional electronic transport in SrTiO$_3$.

The details of sample fabrication are explained elsewhere.[16] Briefly, films were deposited on undoped single crystal SrTiO$_3$ (100) substrates by ablating single crystal undoped and 1 at. % Nb-doped SrTiO$_3$ targets. During growth, the substrates were kept at 1200 °C under a background oxygen pressure of ~ 10$^{-8}$ Torr. The thicknesses of the cap and buffer layers were both 100 nm, while *d* was varied. This cap layer successfully suppressed surface depletion effects.[16,20] After the growth, the samples were postannealed *in-situ* at 900 °C under 10$^{-2}$ Torr oxygen partial pressure for 30 minutes to refill the oxygen vacancies created by the extreme reducing growth conditions. The total thickness including buffer, doped, and cap layers was measured by a stylus profiler, and $d$ was estimated by the pulse number, assuming the same deposition rate for all layers. For transport measurements, the samples were cut into a rectangular shape of ≈ 0.5 mm × 3 mm, and contacts to the dopant layer were made using Al wire and ultrasonic wirebonding. Transport properties were made in a helium-4 cryostat with temperatures down to 2 K, and magnetic fields up to 14 T.

Figure 1(b) shows the surface morphology of a typical sample, measured by atomic force microscopy, showing a clear step-and-terrace structure with a step height of one unit cell (≈ 0.4 nm), which assures negligible interface roughness at the doped layers. Figure 2 shows the temperature dependence of the sheet resistance ($R$) for the δ-doped samples with a variety of doped layer thicknesses. This figure indicates that thick samples are metallic down to 2 K, while thin samples with higher $R$ show an upturn at low temperature, which is indicative of localization. Hall measurements in the low field limit showed that the sheet carrier density ($n_{2D}$) is almost temperature independent (not shown), and the variation of conductivity is solely driven by electron mobility ($\mu$), given by $\mu = 1/n_{2D} e R$ ($e$ is the elementary electronic charge and $\mu$ is the electron mobility).

Next we focus on the low-temperature transport properties. Figures 3(a) and 3(b) show $n_{2D}$ and $\mu$ estimated from the Hall effect for magnetic fields between ±1 T, as a function of $d$ at 2 K, indicating that $n_{2D}$ is proportional to $d$ and consistent with the nominal values indicated by the dashed lines in the figure. The electron mobility is, on the contrary, initially enhanced as $d$ decreases, reaching the maximum value of ~ 1,500 cm$^2$ V$^{-1}$ s$^{-1}$. It is remarkable that this value is about six times larger than the bulk values for the equivalent doping as indicated by the arrow in Fig. 3(b). By further decreasing $d$, both $n_{2D}$ and $\mu$ are sharply reduced below a critical thickness of ~ 3 nm, eventually leading to a completely insulating state.

For a deeper understanding of the electronic structure, we need to estimate the out-of-plane carrier distribution. Previously, we have confirmed negligible diffusion of Nb dopant in SrTiO$_3$ thin films, using secondary ion mass spectroscopy.[16] The carriers themselves, however, can extend beyond the doped region. To estimate the carrier distribution, we measured Seebeck coefficient ($S$) and extracted the three-dimensional carrier density ($n_{3D}$) using[21]

$$S = -k_B/e[(r+2)F_{r+1}(\xi)/(r+1)F_r(\xi) - \xi], \qquad (1)$$

where $k_B$ is the Boltzmann constant, $\xi$ is the chemical potential, $r$ is the scattering parameter of relaxation time, and $F$ is the Fermi integral. Here, $F$ and $n_{3D}$ are given by

$$F_r(\xi) = \int_0^\infty x^r/(1+e^{x-\xi})dx \qquad (2)$$

and

$$n_{3D} = 4\pi(2m^*k_B T/h^2)^{3/2} F_{1/2}(\xi), \qquad (3)$$

respectively. This analysis is valid for the thickness range studied, which is larger than the ~ 3 nm crossover below which thermopower enhancement in electron doped SrTiO$_3$ has been observed.[22] Figure 3(c) compares $n_{3D}$ estimated from the thermopower measurements and $n_{2D}/d$ estimated from the low-field Hall measurements. This shows good agreement between the two data sets within experimental error. The fact that the nominal growth thickness is in good agreement with the measured thickness was also previously concluded by measuring the anisotropy of superconducting upper critical field.[6] All of these facts support a narrow confinement of carriers within the doped layer.

This observation is rather surprising because the large dielectric constant of SrTiO$_3$ ($\varepsilon \approx 20000\ \varepsilon_0$ at low temperature[23]) would be expected to form an extremely weak confinement fixed-charge potential (Fig. 1(a)). However, the dielectric constant of SrTiO$_3$ is known to be strongly suppressed by moderate electric fields,[12] which reflects the incipient ferroelectric character of SrTiO$_3$.[23] Therefore, the observed narrow carrier confinement can be ascribed to the electric field produced by the charged impurities, together with the suppressed dielectric constant, as was suggested by previous subband calculations.[6]

We now discuss the origin of the mobility enhancement in the thin limit. First we note that $n_{3D}$ is slightly reduced in the thin limit both from the Hall and thermoelectric measurements as shown in Fig. 3(c). Following the bulk phase diagram, a mobility enhancement would be expected if $n_{3D}$ is decreased due to the partial localization of the doped carriers. However, we find that $\mu$ is already enhanced in a regime where $n_{3D}$ does not show any significant reduction from the bulk value, indicating that the mobility enhancement cannot be interpreted solely in terms of the bulk mobility properties.

Similar enhancement of $\mu$ has been commonly observed in multisubband δ-doped GaAs structures.[19] and attributed to energy level quantization and a broadening of wavefunctions in the confined potential. In our δ-doped systems, Hall resistance measurements up to 14 T indeed showed significant nonlinearity when $d$ was reduced below 40 nm, as shown in Fig. 4. This is associated with multiple carrier conduction.[24] This crossover to the non-linear Hall regime occurs at a thickness that roughly corresponds to the value below which $\mu$ shows significant enhancement suggesting that a nontrivial number of the conduction electrons lie outside of the doped layer in the thin limit, giving a parallel conduction path of higher mobility electrons in the surrounding, undoped clean SrTiO$_3$.

In summary, we fabricated δ-doped SrTiO$_3$ structures with various doped layer thicknesses using an improved growth technique of pulsed laser deposition. The transport measurement as well as thermoelectric measurements indicated that the carriers are controllably doped in a much narrower region than expected from the large dielectric constant of SrTiO$_3$. This sharp confinement can be understood by taking into account the strong suppression of dielectric constant of SrTiO$_3$. Reflecting this strong confinement, the low-temperature electron mobility is enhanced by the quantum-mechanical broadening of electron wavefunctions as the doped layer became thinner. Based on these facts, δ-doping shows great potential to be an ideal way to fabricate controlled structures in order to realize two-dimensional high-mobility electron gases in SrTiO$_3$.

This work is supported in part by the Department of Energy, Office of Basic Energy Sciences, Division of Materials Sciences and Engineering, under contract DE‐AC02‐76SF00515.

**Figure Captions:**

**Fig. 1:** (Color online) (a) Schematic diagram of the δ-doped SrTiO$_3$ structure showing the Nb-doped layer embedded in undoped layers, together with the electron density distribution *n* associated with the confining potential *V*. (b) Surface morphology measured by atomic force microscopy.

**Fig. 2:** (Color online) Temperature dependence of the sheet resistance of the δ-doped SrTiO$_3$ samples with a variety of doped layer thicknesses.

**Fig. 3:** (a) Sheet carrier density and (b) electron mobility as a function of doped layer thickness at 2 K, estimated from the Hall coefficient in a magnetic-field range of $\pm 1$ T. (c) Three-dimensional carrier density as a function of doped layer thickness, estimated from the low-field Hall coefficient at 2 K (filled circle) and thermopower measurements at room temperature (open circle). The dotted curves in (a) and (c) show nominal values, while in (b) is guide to the eye. The bulk mobility is indicated by the arrow in (b).

**Fig. 4:** (Color online) (a) Hall resistance versus magnetic field at 2 K, normalized by the zero-field Hall coefficients for a variety of doped layer thicknesses.

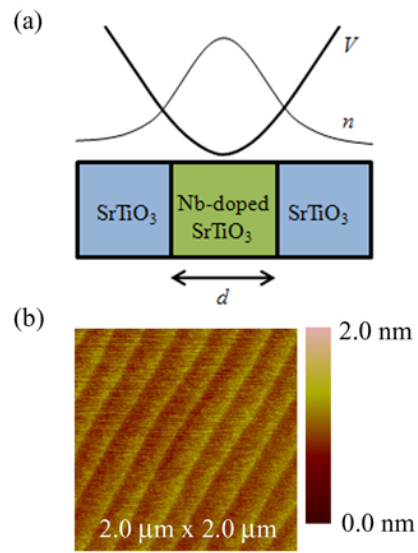

Fig. 1. Y. Kozuka et al.

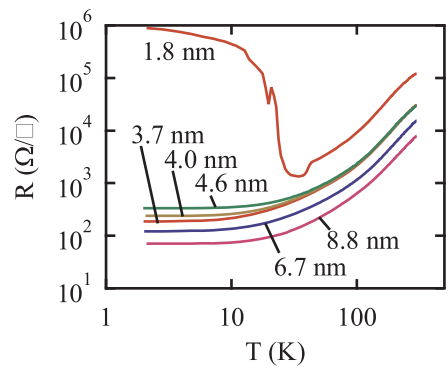

Fig. 2. Y. Kozuka et al.

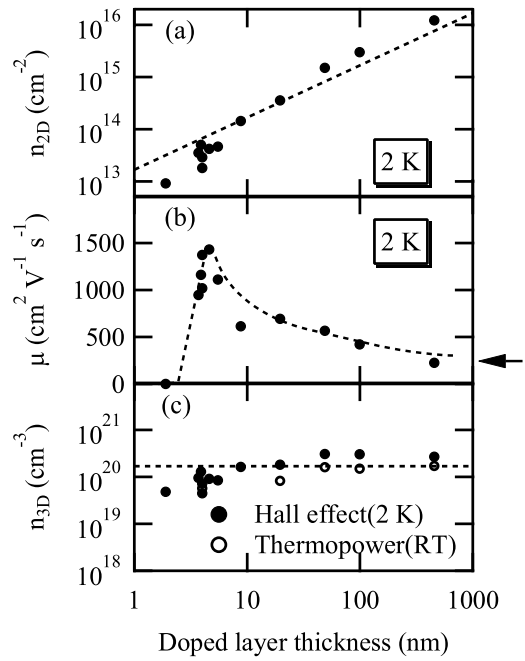

Fig. 3. Y. Kozuka et al.

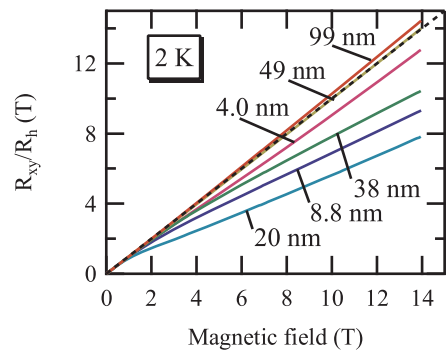

Fig. 4. Y. Kozuka et al.